\documentstyle[12pt]{article}

\def\ifm{\ifmmode}
\def\go{\ifm \rightarrow \else $\rightarrow$\fi}
\def\shat{\ifm \hat{s} \else $\hat{s}$\fi}
\def\gam{\ifm \gamma \else $\gamma$\fi}
\def\pt{\ifm p_{\rm T} \   \else $p_{\rm T}$\ \fi}
\def\ptm{\ifm p_{\rm T\min} \  \else $p_{\rm T\min}$\ \fi}
\def\GeV{\rm GeV}
\newcommand{\beq}{\begin{eqnarray}}
\newcommand{\eeq}{\end{eqnarray}}

\begin{document}
\renewcommand{\thesection}{\arabic{section}.}
\begin{flushright}
FSU--HEP--930609\\
hep-ph/9307238\\
June 1993
\end{flushright}
\begin{center}
{\large\bf Measuring the Longitudinally Polarized Proton Gluon
Distribution using Photoproduction Processes}

\vspace{.7in}

{S.~Keller and J.~F.~Owens}

\vspace{.55in}

{\it Department of Physics, B-159\\
Florida State University\\
Tallahassee, Florida 32306}
\end{center}
\vfil
\begin{abstract}\normalsize\baselineskip24pt

Little information is known about the polarization of gluons inside  a
longitudinally polarized proton. We investigate the sensitivity of
photoproduction experiments with both beam and target longitudinally
polarized to the polarization of the gluon distribution in the proton.
We study the photoproduction of jets and heavy quarks and conclude that
they are both sensitive to the gluon polarization.

\end{abstract}
\vfil

\newpage
\section{Introduction}

Since the so-called EMC spin crisis has emerged\cite{EMC88,EMC89}, much
theoretical work has been done\cite{Bas92}.  Central to the debate is the size
of the gluon polarization  inside a longitudinally polarized proton.   The only
information available about  the gluon polarization is given through higher
order corrections to spin-dependent structure functions or through its effect
on the evolution of polarized quark distribution functions.   However by
considering reactions for which  the gluon plays a role at leading order, the
sensitivity to its polarization will be increased.  Many ways to measure the
gluon polarization in this more direct fashion have been suggested in polarized
deep inelastic  scattering\cite{Car88,Glu88,Alt89} and polarized hadron-hadron
interactions\cite{Ber89}. In this paper, the sensitivity to the gluon
polarization is studied in photoproduction experiments where both the photon
and the proton are longitudinally polarized.  The production of both jets and
heavy quarks is considered.    Although these processes have been  studied
before\cite{Kun89,Bab79}, they are reconsidered here in more  detail and  in
light of a possible large gluon polarization.

In Section 2, leading order sets of polarized distribution functions
compatible with the EMC result and the work of Bodwin and  Qiu\cite{Bod90} are
presented.  Different amounts of gluon polarization are considered. As is well
known, the photon distribution functions play an important role in the
photoproduction of jets.  The polarized distribution functions of the photon
developed in Ref.~\ref{Has81} are used.  Their important features are
summarized  in Section 3. In Sections 4 and 5, the results for the
photoproduction of jets and heavy quarks are discussed, respectively.
Finally, in Section 6 the conclusions are reviewed.

\section{Distribution functions of the proton}

\indent

The relevant quantities are the helicity difference distribution functions:
\beq
\nonumber \\
\Delta q_i(x_p,Q^2)=q_i^+(x_p,Q^2) - q_i^-(x_p,Q^2),\\
\nonumber
\eeq
where $q_i^+(x_p,Q^2) (q_i^-(x_p,Q^2))$ is the probability density for a quark
of flavor i and momentum fraction $x_p$ to have a  helicity of the same
(opposite) sign as the helicity of the proton. The same expression applies for
the anti-quarks and gluon. $Q^2$ is the QCD evolution scale. Assuming an SU(3)
symmetric sea, the following parametrizations are used at $Q^2=Q_0^2=4\
\GeV^2$:
\beq
\nonumber \\
x_p \Delta u_v(x_p,Q_0^2)&=&N_{u_v} x_p^{a_{u_v}}
(1-x_p)^{b_{u_v}}(1+\gam_{u_v} x_p)
\nonumber \\
x_p \Delta d_v(x_p,Q_0^2)&=&N_{d_v} x_p^{a_{d_v}} (1-x_p)^{b_{d_v}},
\nonumber \\
x_p \Delta s(x_p,Q_0^2)&=&N_s x_p^{a_s} (1-x_p)^{b_s},
\nonumber \\
x_p \Delta c(x_p,Q_0^2)&=& 0,\\
\nonumber
\label{eq:dis}
\eeq
where
\beq
\nonumber \\
N_{u_v}&=&\Delta u_v(Q_0^2)/(\beta (a_{u_v},b_{u_v}+1)+\gamma_{u_v} \beta
(a_{u_v}+1,b_{u_v}+1)), \nonumber \\
N_{d_v}&=&\Delta d_v(Q_0^2)/\beta (a_{d_v},b_{d_v}+1), \nonumber \\
N_{s}&=&\Delta s(Q_0^2)/\beta (a_{s},b_{s}+1),\\
\nonumber
\eeq
$\beta(x,y)$ is the Euler beta function,
and $\Delta q_i(Q_0^2)$ is the first moment of $\Delta q_i(x_p,Q_0^2)$:
\beq
\nonumber \\
\Delta q_i(Q_0^2)=\int_{0}^{1} \Delta q_i(x_p,Q_0^2) dx_p.\\
\nonumber
\eeq
All the quarks are assumed to be massless. Not enough experimental information
is available to fit all  the parameters, so that some phenomenological input
has to be used. A strong constraint is provided by the requirement that the
helicity difference distribution functions must be smaller than the
corresponding unpolarized (helicity--sum)  distribution functions.  The
parameters that are used satisfy this requirement with respect to the set
DO1.1\cite{Owe91}.  The small $x_p$ behavior is fixed,  following Altarelli and
Stirling\cite{Alt89}, by taking  $a_{u_v}=a_{d_v}=0.8$ and $a_s=0.7$.  The high
$x_p$ behavior is assumed to be similar to the behavior of the unpolarized
distribution functions: $b_{u_v}=3.7$,  $b_{d_v}=4.7$, and $b_s=8$. Once an
SU(3) symmetric sea is assumed, the normalization of the valence quarks are
fixed by the Bjorken sum rule and results from hyperon beta decay\footnote{
Actually only the first moments need to be SU(3) symmetric for this to be
true.}.  Following Ref.~\ref{EMC88}, $\Delta u_v(Q_0^2)=0.97$ and $\Delta
d_v(Q_0^2)=-0.28$ are used.   The sea normalization can be fixed with the EMC
result on the spin dependent structure function:
\beq
\nonumber \\
g_1^p(x_p,Q^2)={1 \over 2} \sum_i e_i^2 (\Delta q_i(x_p,Q^2)+\Delta \bar
q_i(x_p,Q^2))\\
\nonumber
\eeq
where $e_i$ is the charge of $q_i$.  Several groups have suggested that the
difference between the Ellis-Jaffe~\cite{Ell74} sum rule and the  EMC result
for the first moment of $g_1^p(x_p,Q^2)$  could be resolved by a larger higher
\sloppypar
order  correction due to a large gluon polarization\cite{Car88,Alt88} or by a
large  sea contribution\cite{Glu88,Ell87}(the sea contribution was set to zero
in the Ellis-Jaffe sum rule).  Several sets of helicity difference
\endsloppypar
distribution functions have been developed based on these
ideas\cite{Alt89,Bou90}. However Bodwin and Qiu\cite{Bod90} showed that a hard
gluonic  contribution to the  first moment of $g_1^p(x_p,Q^2)$ vanishes if an
appropriate regularization scheme is used.   Thus, the only way to resolve the
discrepancy between the Ellis--Jaffe sum rule and the EMC result seems to be to
assume a ``large'' negative sea  contribution.  This point of view is adopted
here and $\Delta s(Q_0)=-0.11$ is taken.  When the first moment of
$g_1^p(x_p,Q^2)$ is estimated, the data are assumed to be $Q^2$ independent.
The EMC result would prefer a somewhat larger $a_s$ and larger negatively
polarized sea when the $x_p$ and $Q^2$ dependence are taken into account.
However, as mentioned earlier, the unpolarized distribution functions give a
strong constraint and limit $\Delta s(Q_0^2)$ to about the value it was
assigned. Note that with these values for the parameters, the spin carried by
the  quarks is close to zero and,  as was noted in Ref.~\ref{EMC88}, this still
needs to be explained.  The last parameter $\gamma_{u_v}$ can be fit to
reproduce  the $x_p$ and $Q^2$  dependence of $g_1^p(x_p,Q^2)$ measured by the
EMC collaboration as well as possible.  The result is $\gam_{u_v}$=2.54.
Because of the large sea contribution, there is no need for a large  higher
order correction, and a leading order  analysis is suitable.  This is
consistent with the fact that later on in this paper leading order expressions
are used. With this approach there is no experimental constraint on the
helicity difference gluon distribution functions \footnote{The constraint from
evolution is small. The situation is similar  to a leading order analysis of
$F_2^p$ in the unpolarized case.}. The following parametrization is used:
\beq
\nonumber \\
x_p \Delta g(x_p,Q_0^2)&=&N_g x_p^{0.6} (1-x_p)^{8},\\
\nonumber
\eeq
where
\beq
\nonumber \\
N_g&=&\Delta g(Q_0^2)/\beta (0.6,1.8).\\
\nonumber
\eeq
The value of the parameters are such that $\Delta g(x_p) \leq g(x_p)$  and that
$\Delta g(Q_0^2)$ $\sim 5 $ is allowed.   Three cases are considered: $\Delta
g(Q_0^2)=.5, 3.0,$ and $5.7$, respectively labeled set 1, 2, and 3.   In Fig.1
both the helicity difference and unpolarized quark and gluon  distribution
functions are shown at $Q_0^2=4\ \GeV$ for the three sets. The distribution can
be obtained at another $Q^2$ by evolving them with  spin-dependant
Altarelli-Parisi equations\cite{Alt77}.   In Fig.2 the EMC results, with
statistical and systematic uncertainties added in quadrature are compared to
the results of set 1 and 3, calculated at the  appropriate $Q^2$.  The
agreement between the data and the sets is good  considering the fact that the
sets are not the result of a fit of all the parameters.  The difference between
the two sets is due to the difference in the evolution of the sea quark
helicity difference distribution functions.

In this section, three sets of helicity difference distribution functions were
developed.  At $Q^2=Q_0^2$, they differ only in the size of the gluon
polarization.  As was pointed out, even the quark distribution functions are
not well constrained yet.  For the purpose of this paper, it is assumed that
the quark distribution functions are fixed and the emphasis is on the
sensitivity to the gluon helicity difference distribution function.

\section{Distribution functions of the photon}

As in the unpolarized case\cite{Wit77},  an``asymptotic'' solution can be
derived from the
spin-dependent Altarelli-Parisi evolution equations in the limit
of large momentum fraction ($x_\gam$) and $Q^2$.  This is due to the direct
coupling of the photon to quarks. In that asymptotic limit, the $Q^2$
dependence
of the helicity difference  distribution functions can be factorized:
\beq
\nonumber \\
\Delta q_i(x_\gam,Q^2) &=&{\alpha \over 2\pi} ln({Q^2 \over \Lambda^2})
\Delta h_i(x_\gam),
\nonumber \\
\Delta g(x_\gam,Q^2) &=&{\alpha \over 2\pi} ln({Q^2 \over \Lambda^2})
\Delta h_g(x_\gam),\\
\nonumber
\label{Eq:delq}
\eeq
where $\alpha$ is the QED coupling constant and $\Lambda$ the QCD constant.
The $\Delta h$ functions depend only on $x_\gam$.   The expression in
Eq.~\ref{Eq:delq} can be introduced in the spin-dependent Altarelli-Parisi
evolution equations and the remaining $x_\gam$ dependent equations for the
$\Delta h$ can be solved numerically. This was carried out in Ref.~\ref{Has81}
where a parametrization of the  solution is  provided~\footnote{At high
$x_\gam$ the parametrization used for the quarks is bigger than the numerical
result and the unpolarized  distribution functions. Here, to correct for this,
$\Delta q(x_\gam)=q(x_\gam)$ is used for $x_\gam > 0.95$.}. Four flavors and
massless quarks were assumed. A solution valid at lower $Q^2$ and $x_\gam$
would require input distributions functions at a fixed $Q^2$.  Unfortunately,
there are no experimental data to fix the parameters of these distribution
functions, so the asymptotic solution is utilized for the entire $Q^2$ range
under
consideration. The same approach can be used in the unpolarized case and the
parametrization of  Ref.~\ref{DO} is used here.  In Fig. 3 both the polarized
and unpolarized distribution functions  are shown for up-type and down-type
quarks.  The distribution functions for the quarks are harder than in the
proton case, as can be seen by comparing Fig. 1 and 3.  This will favor
configurations with high $x_\gam$ and low $x_p$.   The polarized distribution
function of the gluon is also shown in Fig. 3 even  though at the energy
considered in this paper, it doesn't play any role. Notice that at high
$x_\gam$ the helicity difference distribution functions of the quarks are
positive.

\section{Two--jet production}
\indent

Our goal is to study the sensitivity of the photoproduction of two jets  to the
gluon polarization.  Both the photon and the proton are longitudinally
polarized. The photoproduction of two jets receive contributions from two
classes  of subprocesses.  In the first class of subprocesses the photon
interacts directly with the constituents of the proton (the ``direct''
contribution).  In the second class, the photon interacts through its
distribution functions (the ``resolved'' contribution).   At leading order in
perturbative QCD, the integrated  cross section for the photoproduction of two
jets  is given by the following expression:
\beq
\nonumber \\
\displaystyle
\sigma_{\gam p \go 2jets}(s)&  = & \sum_{a,b}
\int_{\tau}^{1}
dx_\gamma\, f_{a/\gamma}(x_\gamma,Q^2)
\int_{\tau_\gam}^{1}
dx_p \, f_{b/p}(x_p,Q^2)\,
\hat \sigma_{a b \go 2jets}(\shat)
\nonumber \\
& & \\
\nonumber
\label{eq:sig}
\eeq
where $s=2E_\gamma M_p$ is the square of the total energy available and
$\shat=x_\gamma x_p s$ is  the square of the center of mass energy of the
subprocesses.   $f_{a/\gamma}(f_{b/p})$ is the distribution function of parton
a(b) in  the photon(proton).  In the direct contribution  case,
$f_{a/\gamma}(x_\gamma)$ is replaced by $\delta(1-x_\gamma)$.   The choice of
the scale $Q^2$ bears the usual ambiguity,  $Q^2=p^2_T/4$ is used. The sum is
over all the possible subprocesses. The direct contribution is composed of two
subprocesses: $\gam q \rightarrow g q$ and $\gam g \rightarrow q \bar q$. The
resolved contribution is composed of eight subprocesses. The dominant
subprocesses at the energy considered in this paper are  $qq'\rightarrow qq'$
and $gq\rightarrow gq$. The lower limits of integration are given by the
following expressions:
\beq
\nonumber \\
\tau&=&4 p^2_{Tmin}\over s \nonumber \\
\tau_\gam&=&4 p^2_{Tmin}\over {s x_\gamma}. \\
\nonumber
\label{eq:tau}
\eeq
where \ptm \  is the minimum \pt \  of the jets. The matrix elements necessary
to calculate the subprocess cross sections, $\hat \sigma$, can be found in
Ref.~\ref{Owe80} and \ref{Bar87}. The integrated cross sections are presented
in Table~\ref{tab:jsig}, at $E_\gamma=200\ \GeV$ and $400\ \GeV$, and $\ptm =
3$ and $5\ \GeV$.  $E_\gam=200\ \GeV$ corresponds to the average value for
$E_\gam$ of  present unpolarized  experiments and  $400\ \GeV$ is about the
upper limit.   $\ptm=3\ \GeV$ is the lowest value at which jets have been
observed in fixed target  experiments\cite{Cor92}.   $\ptm =5\ \GeV$ is
presented to show the variation of the different contributions with  \ptm. Both
the direct and the resolved contributions are presented.  The two contributions
are furthermore divided into quark and gluon contributions corresponding to
subprocesses involving a quark or a gluon inside the proton, respectively.  The
unpolarized case has already been study extensively~\cite{Owe80,Bae89}. The
important points are summarized briefly. As can be seen from Eq.~\ref{eq:tau},
both the $x_p$ and $x_\gam$ thresholds  decrease when the energy increases or
the \ptm decreases.  As a result, the relative size of the gluon contribution
(in both the direct and resolved contribution) and  of the resolved
contribution increases whenever the energy is increased or the minimum \pt
decreased.

The integrated helicity difference cross section for the production of two jets
is given by\cite{Bab79}:
\beq
\nonumber \\
\displaystyle
(\sigma^{++}-\sigma^{+-})_{\gam p \go 2jets}(s)&  = & \sum_{a,b}
 \int_{\tau}^{1}
dx_\gamma\, \Delta f_{a/\gamma}(x_\gamma,Q^2)\, \nonumber \\
&& \int_{\tau_\gam}^{1}
dx_p\, \Delta f_{b/p}(x_p,Q^2)\,
\Delta \hat \sigma_{a b \go 2jets}(\shat).\\
\nonumber
\label{eq:asym}
\eeq
The first sign superscript in $\sigma^{++}$ and $\sigma^{+-}$ corresponds to
the helicity of the photon, and the second to the helicity of the proton. The
$\Delta f_{a/\gam}$ and $\Delta f_{b/p} $ are the helicity difference
distribution functions as defined in section 2 and 3. $\Delta \hat\sigma$ is
the helicity difference cross section of the  subprocess.  The matrix elements
for the different subprocesses can be found  in Ref.~\ref{Bab79} and
\ref{Has81}.  The longitudinal asymmetry can now be defined:
\beq
\nonumber \\
A_{ll}={\sigma^{++}-\sigma^{+-}\over \sigma^{++}+\sigma^{+-}}.\\
\nonumber
\label{eq:All}
\eeq
As usual, it is advantageous to consider a ratio because the  theoretical and
experimental uncertainties tend to cancel out.    The  results for the
asymmetry of each of the contributions  are shown in Table~\ref{tab:jasym} for
set 1 and 3 (smallest and largest gluon helicity difference distribution
function),  at the same energies and \ptm \  as in Table~\ref{tab:jsig}. An
important property of the asymmetry is that it is a weighted average of the
asymmetries of each of the contributions.

First, the direct contribution is considered.  The quark contribution is
dominated by the $u_v$ quark because the cross section is  proportional to the
square of the charge of the quark, and the $\Delta u_v$ is the largest of the
quark helicity difference  distribution functions.  Both $\Delta u_v$ and
$\Delta \hat \sigma_{\gam q \rightarrow gq}$ are  positive so that the quark
contribution gives a positive asymmetry.   The difference between set 1 and set
3 for the quark contribution is small; it is  due to the difference in
evolution of the quark helicity difference  distribution functions.   For the
gluon contribution $\Delta g$ is positive and  $\Delta \hat \sigma_{\gam g
\rightarrow q \bar q}$ is negative such that the asymmetry of this contribution
is negative.  As expected, there is a large difference between the results of
the two sets  in this case.  At $E_\gam=200\ \GeV$ and $\ptm=5\ \GeV$  the
difference
between the asymmetry of the two sets for the gluon contribution is about 60\%,
whereas the difference for the total asymmetry is only about 15\%.  The large
difference in the gluon contribution doesn't survive, because the size of the
gluon contribution to the cross section is  relatively small, see
Table~\ref{tab:jsig}.   For the other three cases in Table~\ref{tab:jasym}, the
size of the gluon contribution is bigger and the difference between  the two
sets for the total asymmetry of the direct contribution is about 35 to 50\%.
Notice that it is important to consider the difference between the two sets,
and not just the result of  each set separately.

Second, the resolved contribution is studied. In this case, both the quark and
gluon contributions give a positive asymmetry. As in the direct case, the
difference between the two sets for the quark contribution is small.  The
difference between the two sets for the gluon contribution is not as big as in
the direct case.  This is due to the fact that the asymmetry of the leading
subprocess ($gq \rightarrow gq$) is  not as big, and that the helicity
difference distribution functions of the photon had to be folded in. As a
result, the difference between the two sets for the total asymmetry of the
resolved contribution is at most 15\%.

As can be seen in Table~\ref{tab:jasym}, the difference between the two sets
for the total asymmetry is of the order of 10--15\%.  This  difference is
rather small compared to the range span by the gluon contribution of the direct
contribution. The problem stems from the fact that the gluon contribution is
negative  in the direct case and positive in the resolved case, such that the
two contributions partially cancel each other.  An obvious way to improve upon
this is to separate the direct and resolved contributions, and then  use the
direct contribution to measure the gluon polarization, as it is the most
sensitive contribution.  Eventually, the resolved contribution could be used to
study the quark helicity difference distribution functions of the photon. The
same techniques developed for the unpolarized case can be implemented to
separate the direct and resolved contributions. One way to do this is by
tagging the remnant jet coming from the photon in the resolved
case\cite{Dre89}, another is by a complete  reconstruction of the kinematics
and an appropriate cut on $x_\gamma$~\cite{Fle91}.

More detailed information can be obtained by looking at the longitudinal
asymmetries of the differential cross section.  In Fig. 4 the
$x_p$--distributions are presented at $E_\gam=200\ \GeV$ and $\ptm=3\ \GeV$ for
the three sets developed in Section 2.  In Fig.~4a the unpolarized cross
section is presented for the direct (dashes) resolved (dots), and total (solid)
contributions.   The resolved contribution peaks at higher $x_p$ to compensate
for the lower $x_\gam$. In Fig. 4b,  4c and 4d the asymmetry for the direct,
resolved and total contributions, respectively, are plotted for the different
sets (dashes: set 1, dots: set 2, solid: set 3).  The weighting between the
direct and resolved contribution to form the total contribution is well
apparent.  For each contribution, the largest difference between the three
sets is at low $x_p$. Note that it is possible to reconstruct $x_p$ only if
both jets in the event are measured.  In case this can not be done, the
rapidity distribution of the jets in the  $\gamma$--proton center of mass is
shown in Fig.~5.  Positive rapidity is  in the direction of the incoming
photon.   The largest difference is at larger rapidity that corresponds to the
lowest $x_p$. One could also look at the \pt distributions, and the difference
between the sets is similar.  For comparison, in Fig. 6 and Fig. 7  the same
plots as in Fig. 4 and Fig. 5, respectively, are shown for $E_\gam=400\ \GeV$
and $\ptm =3 \GeV$.

\section{Heavy Quark production}

The formulas presented in Section 4 for the photoproduction of two jets are
also valid for the photoproduction of an heavy quark pair.  The only
modification is the replacement of \ptm by $m_q$, the mass of the heavy quark,
in Eq. 10.   We will consider the production of the charm quark with $m_c=1.5\
\GeV$.    As is well known, the resolved contribution for the photoproduction
of heavy quarks for the energy range considered here  is of the order of a few
percents, and can be neglected.   For the direct contribution there is only one
subprocess at leading order in perturbative QCD: $\gam g \rightarrow  Q \bar
Q$, where $Q$ stands for a heavy quark.  Assuming that there is no spin effect
in the fragmentation of the charm into a D--meson,  the Peterson fragmentation
function can be used~\cite{Pet83}:
\beq
\nonumber \\
D_c^D(z)=N z(1-z)^2/((1-z)^2+\epsilon z)^2\\
\nonumber
\label{eq:Pet}
\eeq
where z is the momentum fraction of the D--meson, and N is taken such that
$D_c^D(z)$ is normalized to 1.  $m_D \sim m_c$ is assumed as in the derivation
of Eq.~\ref{eq:Pet}. The parameter $\epsilon$ is taken at .15. The matrix
elements needed to calculate the asymmetry were evaluated using the method
described in Ref.~\ref{Bar91}. The results for the integrated cross section and
the asymmetry are presented in Table~\ref{tab:qtab}, for sets 1 and 3.   The
difference between the asymmetries of the two sets  is about 12\% at
$E_\gam=200\ \GeV$ and 5\% at $E_\gam=400\ \GeV$. These results are actually
misleading.   In Fig. 8 the $p^2_{T}$--distribution of the D--meson along with
the asymmetry distribution are presented  for $E_\gam=200\ \GeV$.   It is
apparent that at low \pt where the mass terms dominate in the cross section,
the asymmetry is positive, whereas at high \pt where the mass terms are not as
important, the asymmetry is negative (as in the two jet production case).   The
difference between the sets is bigger  than suggested by the integrated
asymmetry.  A better variable in this case to describe the difference between
the sets is given  by an ``absolute'' asymmetry, $|A|_{ll}$, where instead  of
integrating the asymmetry at each phase space point with its sign, the absolute
value at each point is integrated.  The difference between the two sets for the
absolute asymmetry  give a measure of the biggest difference that can be
reached. The results for $|A|_{ll}$ are also shown in  Table~\ref{tab:qtab}.
The difference between the sets is of the order of 20\% at $E_\gam=200\  \GeV$
and 15\% at $E_\gam=400 \GeV$.  The lower energy is favored in this case.
In Fig. 9 and Fig. 10 the rapidity distribution and sum of the rapidity
distributions (in case both D--mesons are reconstructed)  are presented. As it
is unlikely that the kinematics of the whole event can be reconstructed,  the
$x_p$ distribution is not shown.

\section{Conclusions}

We have shown that both jets and heavy quark production in a photoproduction
experiment can lead to a successful measurement of the gluon helicity
difference distribution function.   Considering the total asymmetry, the two
jets and heavy quarks production cases have similar sensitivity.  However,
the small difference in the asymmetries between the different gluon
polarizations (10--20\%) might  be a limiting factor.  The best way to
measure the gluon helicity difference  distribution function is by using two
jet production at low \pt, with separation of direct and resolved
contribution.   The direct contribution has the biggest sensitivity, with
differences in the asymmetries of the order of 35--50\%.
The resolved contribution could be  used to study the quark helicity
difference distribution functions of the photon.

To conclude we emphasize that a direct  measurement of the gluon helicity
difference distribution function is important  as it would clarify some of the
theoretical debate.

\indent

\section*{Acknowledgements}
One of us, S.K., acknowledge a useful discussion with J.~Qiu, and thanks
A.~Stange for providing him with his package to numerically calculate helicity
amplitudes and M. Doncheski for discussions about the polarized distribution
functions.  This research was supported in part by the Texas National Research
Laboratory Commission and by the U.S. Department of Energy under contract
number DE--FG05--87ER40319.

\newpage
{\Large \bf Tables}
\begin{table}[h]
\tenrm
\centering
\begin{tabular}{|c|c|c|c|c|c|c|c|c|}\hline
& & \multicolumn{3}{c|}{direct} &\multicolumn{3}{c|}{
resolved}&tot\\ \cline{3-8}
$E_\gamma$&$\ptm$& q-cont&g-cont&tot   &q-cont&g-cont&tot& \\
(\GeV)    &(\GeV)& ($nb$) &($nb$) &($nb$)&($nb$)&($nb$)&($nb$)&($nb$)\\
\hline\hline
200      &5    &10.2&3.2&13.4&2.5&1.2&3.7&17.1      \\	\hline
200      &3    &103.&110.&213.&91.&110.&201.&414.   \\	\hline
400      &5    &23.7&17.3&41.0&13.9&12.6&26.5&67.5  \\	\hline
400       &3    &129.&217.&346.&231.&371.&602.&973.  \\	\hline
\end{tabular}
\caption
[Integrated cross sections for dijet production.]
{Integrated cross sections for dijet production averaged over
the initial spins for different $E_\gamma$, and \ptm.
}
\label{tab:jsig}
\end{table}

\begin{table}[h]
\tenrm
\centering
\begin{tabular}{|c|c|c|c|c|c|c|c|c|c|}\hline
& & & \multicolumn{3}{c|}{direct} &\multicolumn{3}{c|}{
resolved}&tot\\ \cline{4-9}
set&$E_\gamma$&$\ptm$& q-cont&g-cont&tot &q-cont&g-cont&tot& \\
   &(\GeV)    &(\GeV)& (\%) &(\%) &(\%)&(\%)&(\%)&(\%)&(\%)\\
\hline\hline
1&200 &5    &34.8&-7.9&24.7&10.1&3.4&7.9&21.1	      \\	\hline
3&200 &5    &34.9&-74.5&9.1&10.1&30.1&16.8&10.7     \\	\hline
1&200 &3    &24.9&-8.3&7.8&2.8&2.6&2.7&5.3	      \\	\hline
3&200 &3    &24.9&-93.2&-36.0&2.8&29.&17.2&-10.1    \\	\hline
1&400 &5    &28.4&-8.9&12.7&4.7&3.2&4.0&9.3	      \\	\hline
3&400 &5    &28.9&-89.2&-20.8&4.8&31.6&17.5&-5.7    \\	\hline
1&400 &3    &18.0&-7.5&2.0&.9&1.8&1.4&1.6	      \\	\hline
3&400 &3    &18.2&-84.6&-46.4&.9&20.3&12.9&-8.7     \\	\hline
\end{tabular}
\caption
[Asymmetries for dijet production.]
{Asymmetries for dijet production for different sets of polarized
distribution functions (set 1 and 3 of section 2), $E_\gamma$, and \ptm.
}
\label{tab:jasym}
\end{table}

\begin{table}[h]
\tenrm
\centering
\begin{tabular}{|c|c|c|c|c|}\hline
proton &$E_\gamma$& cross   & $A_{ll}$& $|A|_{ll}$ \\
set    &        & section &         &            \\
       &(\GeV)  & ($nb$)  &(\%)     &(\%)        \\  \hline\hline
1      &200.    &993.    &1.2      &2.2         \\ \hline
3      &200.    &993.    &13.2     &23.6        \\ \hline
1      &400.    &1475.    &.5       &1.5         \\ \hline
3      &400.    &1475.    &5.8      &16.7        \\ \hline
\end{tabular}
\caption
{Integrated cross sections and asymmetries for heavy quarks
production for different sets of polarized
distribution functions (set 1 and 3 of section 2), and $E_\gamma$.
}
\label{tab:qtab}
\end{table}

\newpage
\clearpage
\begin{figure}[h]
\raggedright
\includegraphics{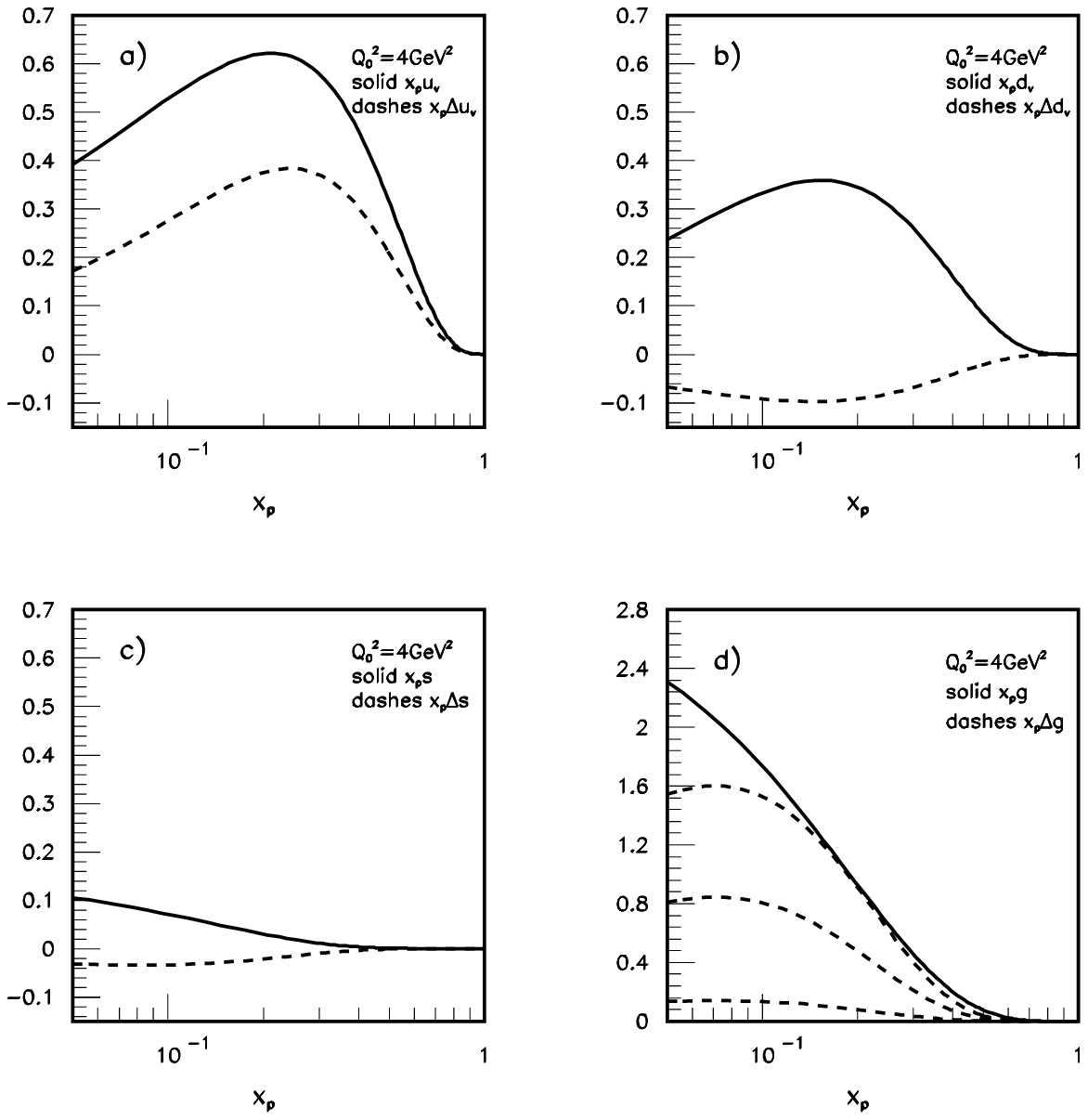}
\caption[fig1]
{Unpolarized distribution functions (set DO1.1, solid) and helicity
difference distribution functions (dashes) of the proton at $Q_0^2=4GeV^2$: a)
valence up quark, b) valence down quark, c) sea quark, and d) gluon.
In Fig. d) the 3 dashed curves correspond from the lowest to the highest
to set 1, 2, and 3.
}
\label{fig:protdist}
\end{figure}

\begin{figure}[h]
\raggedright
\includegraphics{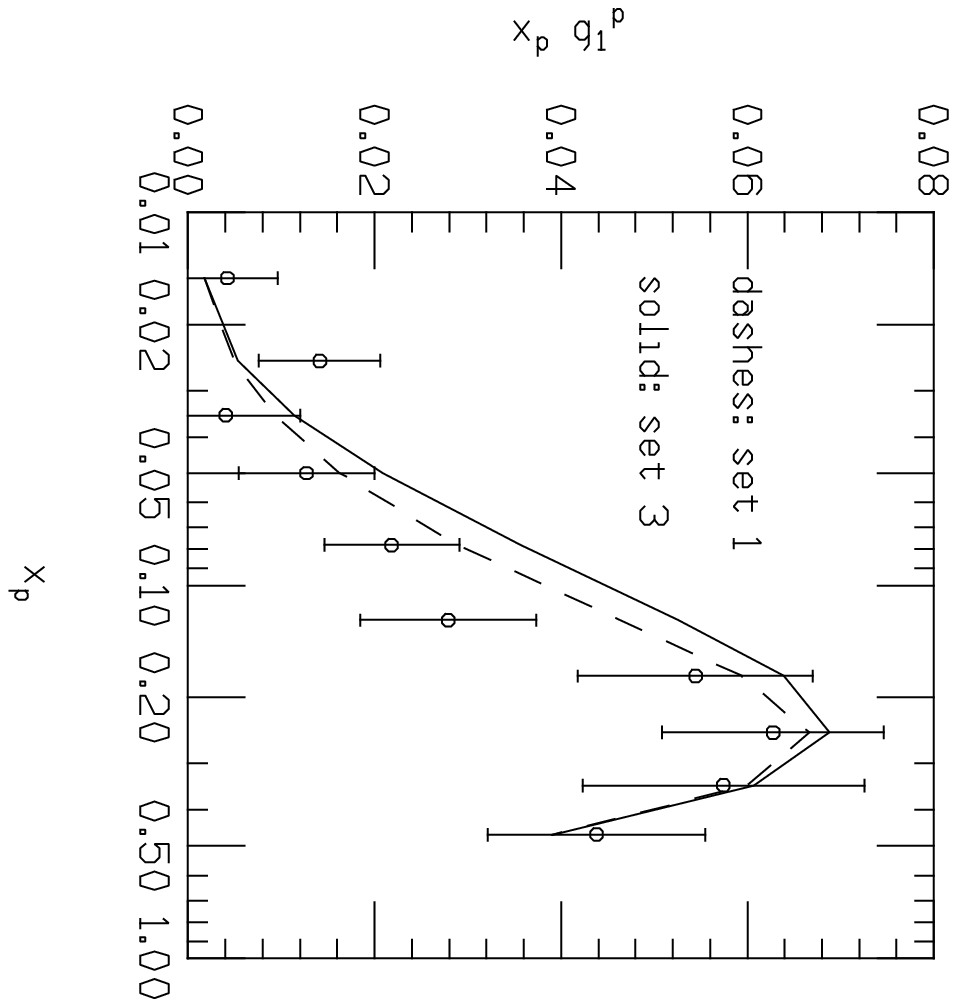}
\caption[fig2]
{EMC result for the spin dependant structure function $g_1^p$
as a function of $x_p$.  Also shown are the results for set 1 (dashes) and
set 3 (solid line), calculated at the appropriate $Q^2$. }
\label{fig:EMC}
\end{figure}

\begin{figure}[h]
\raggedright
\includegraphics{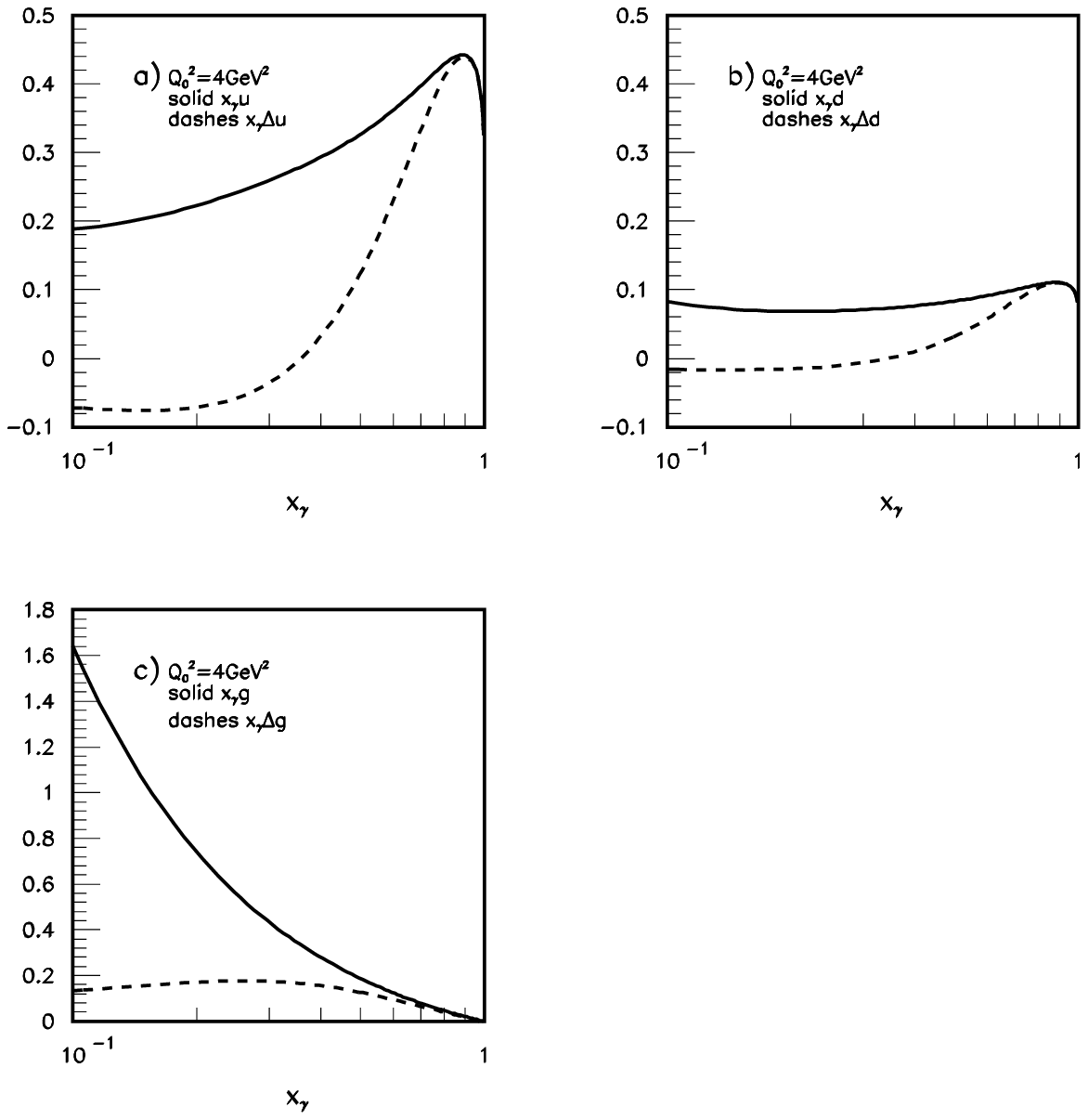}
\caption[fig3]
{Unpolarized distribution functions (Ref.~\ref{DO}, solid) and helicity
difference distribution functions (Ref.~\ref{Has81}, dashes) of the
photon at $Q_0^2=4GeV^2$: a) up type quark, b) down type quark, c) gluon.
}
\label{fig:photdist}
\end{figure}

\begin{figure}[h]
\raggedright
\includegraphics{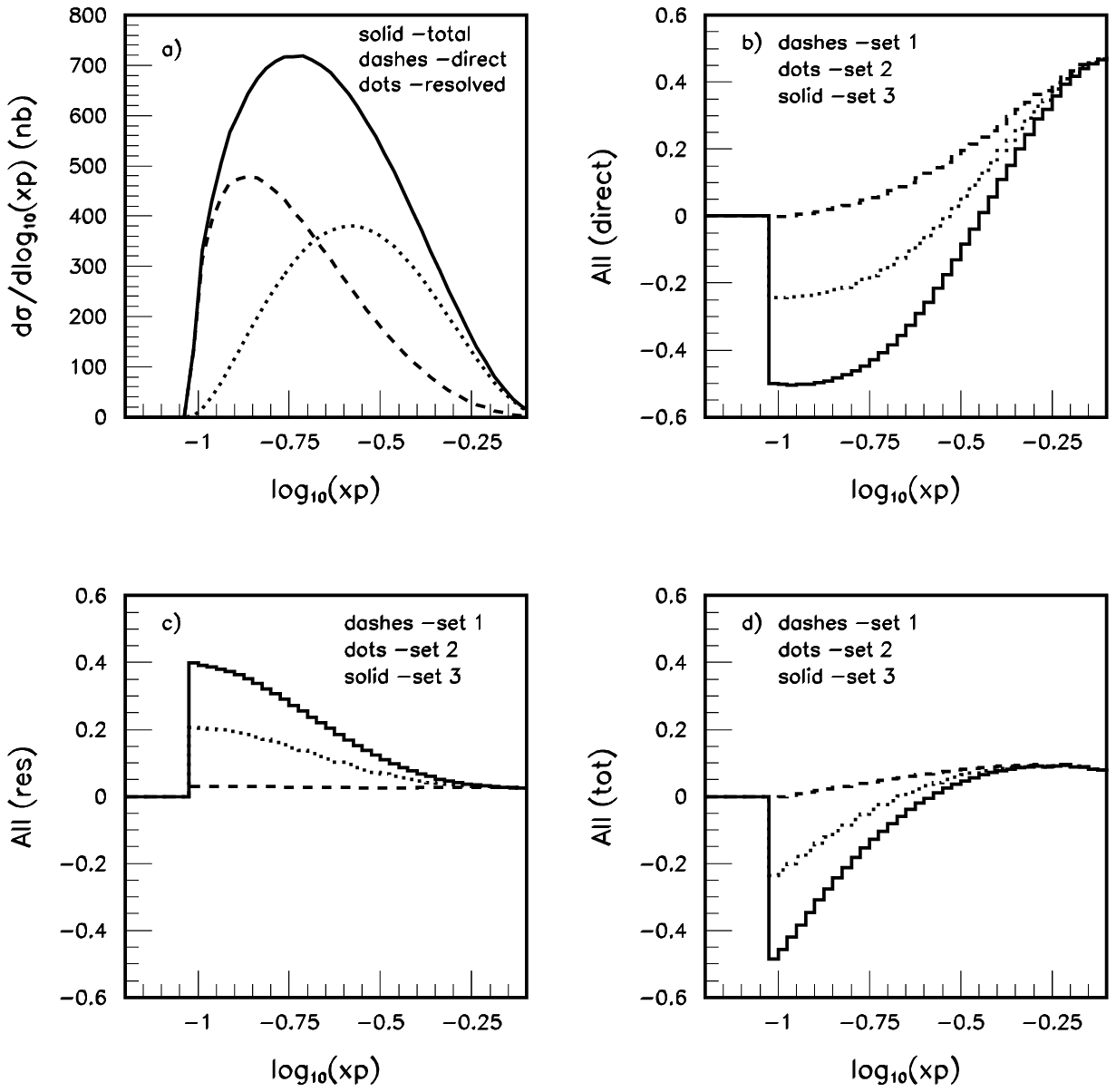}
\caption[fig4]
{Photoproduction of two jets at $E_\gam=200GeV$ and $\ptm=3GeV$.
a) $log_{10}(x_p)$--distribution of the direct (dashes), resolved (dots), and
total (solid) contributions.
b) asymmetry distribution of the direct contribution for set 1 (dashes),
set 2 (dots), and set 3 (solid) as a function of $log_{10}(x_p)$.
c) same as b) for the resolved contribution. d) same as b) for
the total contribution.
}
\label{fig:jxp200}
\end{figure}

\begin{figure}[h]
\includegraphics{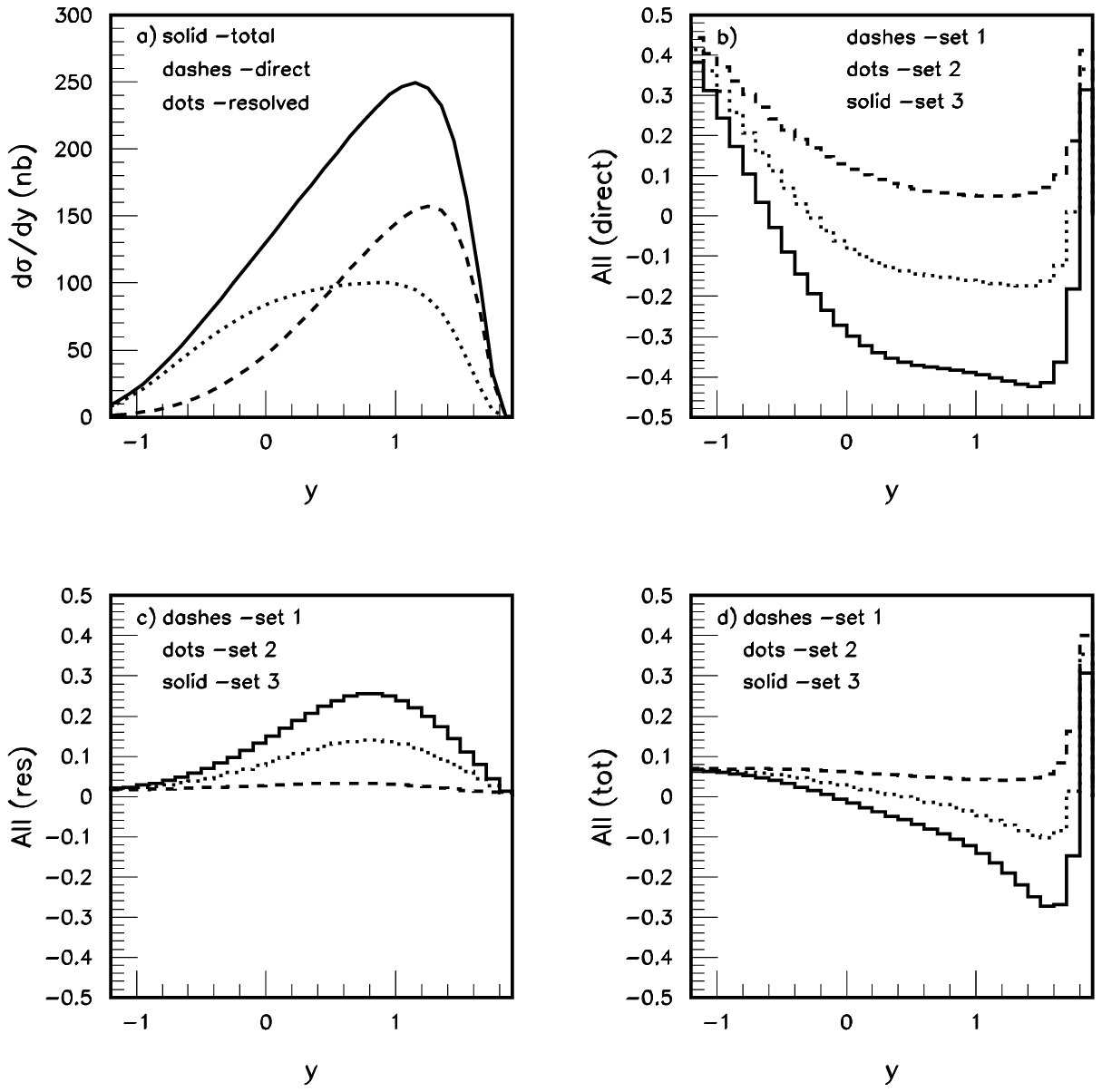}
\caption[fig5]
{Same as in Fig.~\ref{fig:jxp200} for the rapidity distribution of the jets in
the $\gamma$--proton center of mass.
}
\label{fig:jy200}
\end{figure}

\begin{figure}[h]
\raggedright
\includegraphics{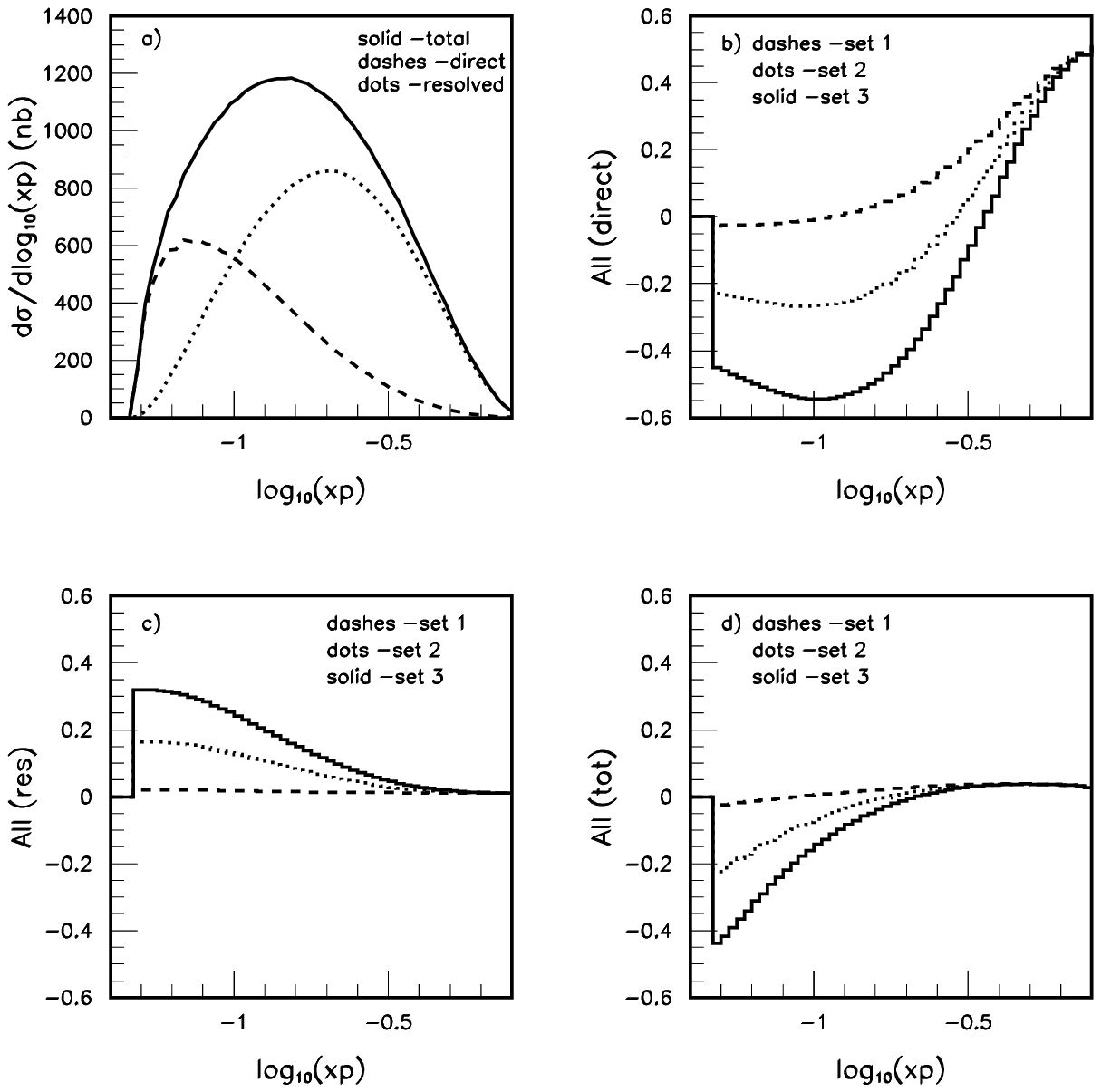}
\caption[fig6]
{Same as in Fig.~\ref{fig:jxp200} at $E_\gam=400\GeV$ and $\ptm=3\GeV$.
}
\label{fig:jxp400}
\end{figure}

\begin{figure}[h]
\raggedright
\includegraphics{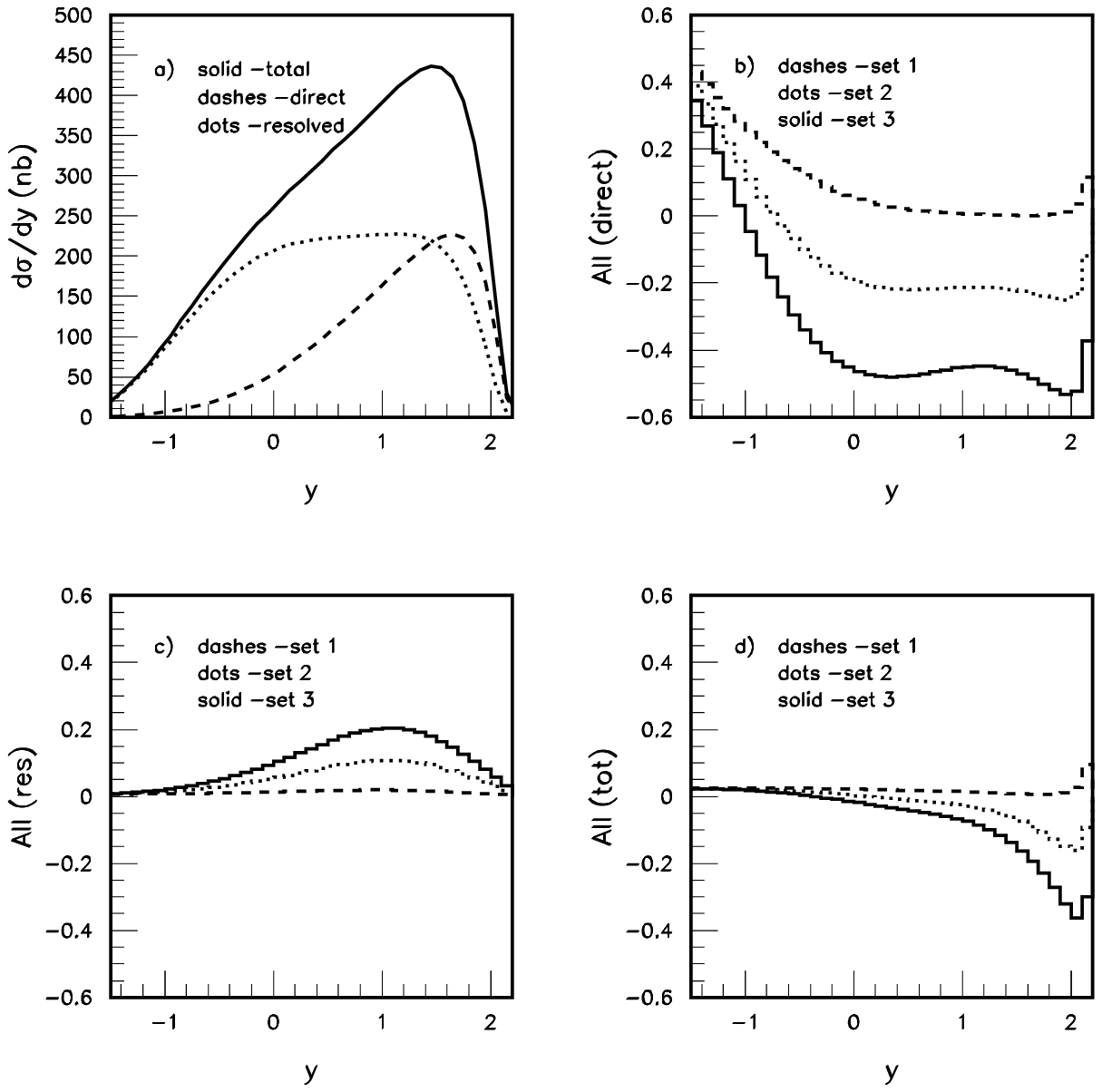}
\caption[fig7]
{Same as Fig.~\ref{fig:jxp400} for the rapidity distribution.
}
\label{fig:jy400}
\end{figure}

\begin{figure}[h]
\raggedright
\includegraphics{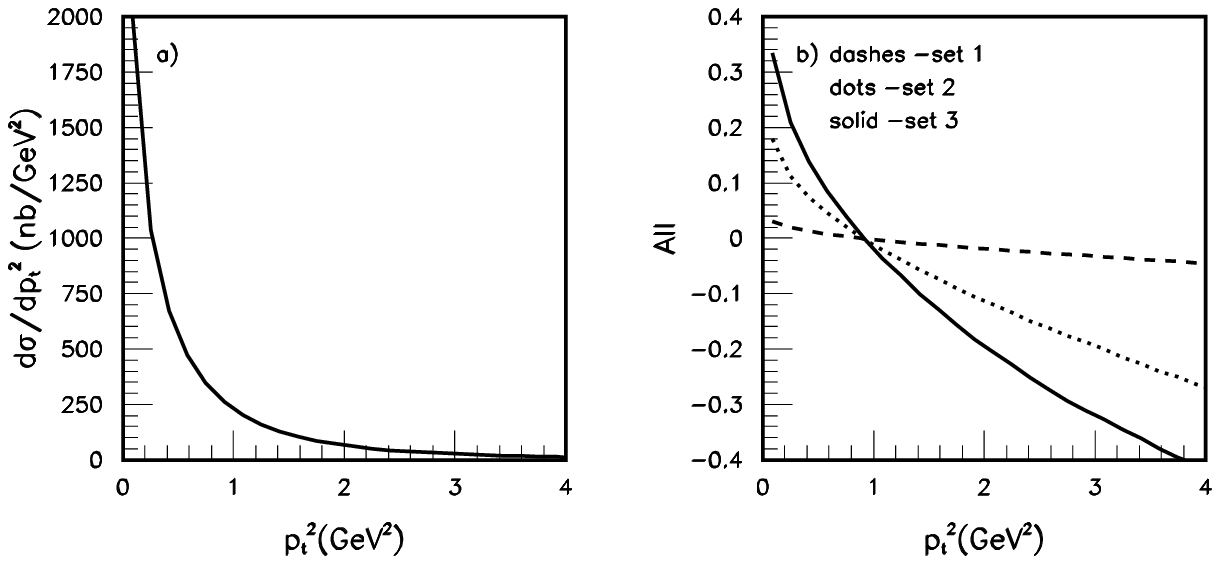}
\caption[fig8]
{Photoproduction of a D--meson pair at $E_\gam=200GeV$.
a) $p_T^2$--distribution. b) Asymmetry as a function of $p_T^2$ for set 1
(dashes), set 2 (dots), and set 3 (solid).
}
\label{fig:qpt200}
\end{figure}

\begin{figure}[h]
\raggedright
\includegraphics{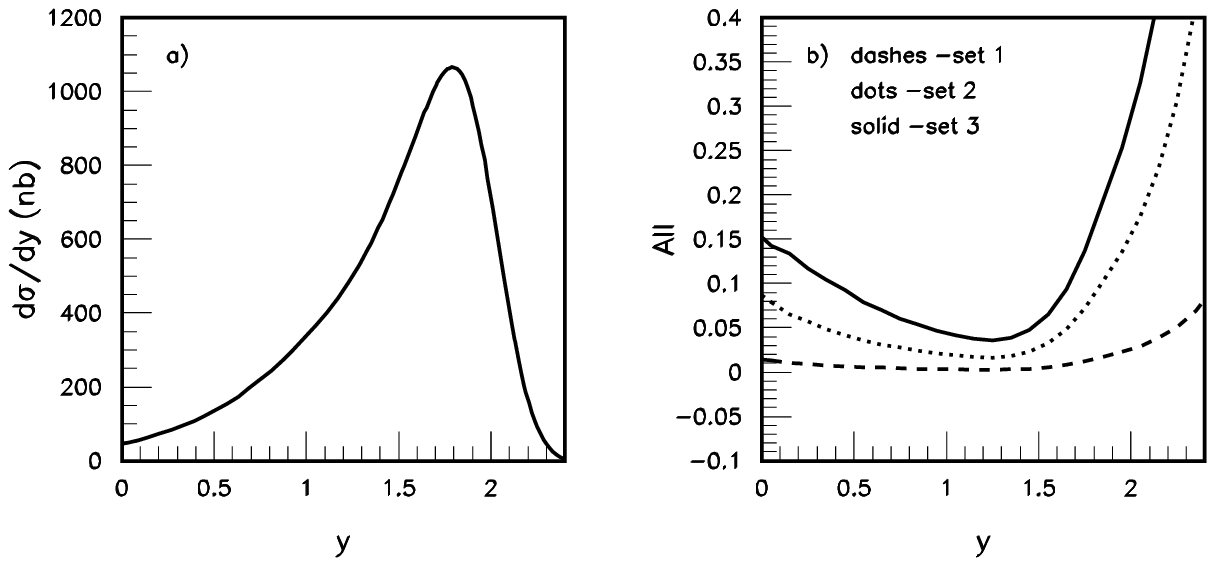}
\caption[fig9]
{Same as in Fig.~\ref{fig:qpt200} for the rapidity distribution of the D--meson
in the $\gamma$--proton center of mass frame.
}
\label{fig:qy200}
\end{figure}

\begin{figure}[h]
\raggedright
\includegraphics{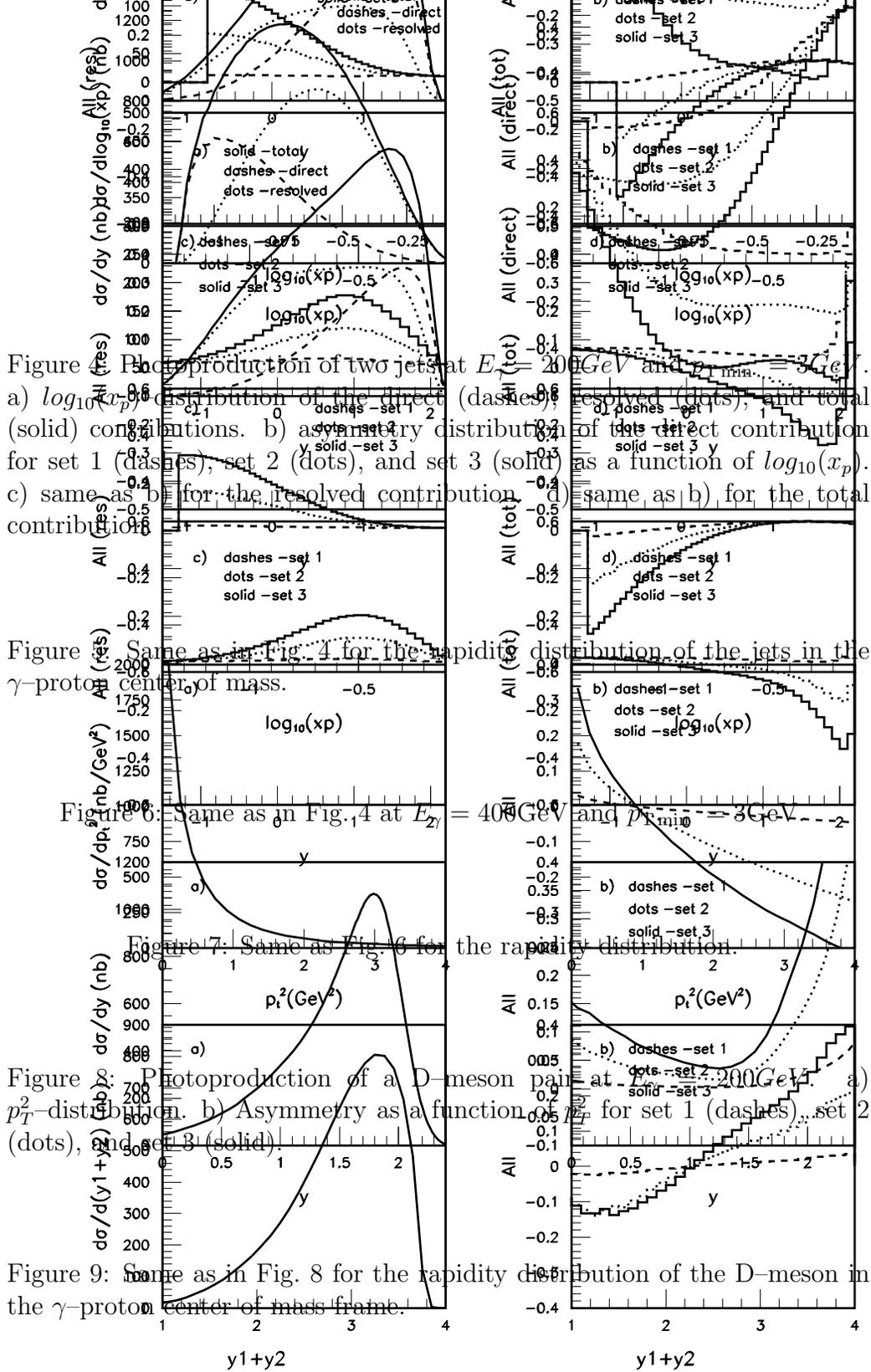}
\caption[fig10]
{Same as in Fig.~\ref{fig:qpt200} for the rapidity--sum distribution.
}
\label{fig:qyy200}
\end{figure}

\end{document}